\documentclass{llncs}

\usepackage{url}
\usepackage{hyperref}
\usepackage[scaled]{beramono}
\usepackage[T1]{fontenc}
\usepackage[usenames,dvipsnames]{xcolor}
\usepackage{amsmath,amssymb,stmaryrd}
\usepackage{graphicx}
\usepackage[utf8]{inputenc}
\usepackage{subcaption}
\usepackage{listings}
\usepackage{alloy}
\usepackage{microtype}
\usepackage{tlatex}
\usepackage{framed}
\usepackage{subcaption}
\usepackage{multirow}

\newcommand{\TLAp}[0]{$\textrm{TLA}^{+}$}

\newcounter{evaluation}

\begin{document}

\title{Alloy meets {\TLAp}: An exploratory study}
\author{Nuno Macedo \and Alcino Cunha}
\institute{{HASLab}  --  High Assurance Software Laboratory\\ INESC TEC \& Universidade do Minho, Braga, Portugal}

\maketitle
\begin{abstract}
Alloy and {\TLAp} are two formal specification languages that are increasingly popular due to their simplicity and flexibility, as well as the effectiveness of their companion model checkers, the Alloy Analyzer and TLC, respectively. Nonetheless, while {\TLAp} focuses on temporal properties, Alloy is better suited to handle structural properties, requiring \emph{ad hoc} mechanisms to reason about temporal properties. Thus, both have limitations in the specification and analysis of systems rich in both static and dynamic properties. This paper explores the pros and cons of these two frameworks when handling this class of systems through the step-by-step modeling, specification and verification of an example.
\end{abstract}

\section{Introduction}
\label{sec:intro}
Software specification and analysis is crucial at early development phases since it encourages the developer to reason about the system and its properties, promoting the detection of design errors. 
This activity generally comprises the \emph{modeling} of the system's acceptable states; the \emph{specification} of the expected behavior of those models; and the actual \emph{verification} of those properties.
Thus, when choosing a specification framework, one must be aware of the class of models and properties expressible in the specification language and manageable by the provided tools. 

The modeling and specification stages require different levels of expressibility from the specification language.
Models are comprised by \emph{structural} properties -- typically first-order logic predicates that specify when the system is considered well-formed -- and \emph{behavioral} properties -- typically action predicates that specify how the system is allowed to evolve. The specification, in contrast, is usually presented in a temporal logic. 
This study focuses precisely on systems rich in both static and dynamic properties.
A paradigmatic example of this class of systems are distributed computing algorithms. Structural properties encode acceptable network topologies and behavioral properties restrict the evolution of their state, while specifications usually take the shape of safety and/or liveness properties. 

Therefore, although a variety of specification frameworks has been proposed, the most successful are characterized by two features: they provide a simple yet expressive and flexible formal language -- allowing the user to specify different classes of systems and properties at different abstraction levels -- and are accompanied by tools that automate their analysis -- providing quick feedback regarding the correctness of the specification.

Alloy~\cite{Jackson:12} is a lightweight formal specification language with an object-oriented flavor, whose companion Analyzer provides support for bounded model checking. Alloy's underlying formalism is \emph{relational logic}, first-order logic enhanced with the transitive closure operation, that renders the definition of structural properties extremely simple. Alloy is inherently static, thus the definition of dynamic properties usually relies on well-known idioms that have emerged thanks to the language's flexibility. Yet, such \emph{ad hoc} solutions are error-prone and force the developer to be concerned with particularities of the idiom rather than with those of the actual system. As a consequence, considerable research has been dedicated to enhance Alloy with dynamic behavior~\cite{FriasGPA:05,ChangJ:06,NearJ:10,VakiliD:12,Cunha:14}. The main drawback of these approaches is that they compromise the flexibility that characterizes Alloy, introducing syntactic extensions that force users into specific idioms.

In contrast, temporal model checkers focus on the analysis of dynamic systems. Among the most successful formalisms is the \emph{temporal logic of actions} (TLA)~\cite{Lamport:02}, a variant of temporal logic that relies on the notion of action to model dynamism. Actions are essentially predicates that relate two consecutive states, and can be used to model the steps that allow the system to evolve. Specification properties are instead defined in a restricted temporal logic. The {\TLAp} specification language is based in this formalism and is accompanied by a useful toolset that includes TLC, a model checker that has proven effective on the verification of complex systems. Although {\TLAp} does support first-order logic properties, these are not the main focus of {\TLAp}, giving rise to some limitations.

Both these frameworks are being increasingly adopted and have been successful applied to complex systems (e.g.~\cite{BagheriKMJ:15} and~\cite{Newcombe:14}). However, although their popularity arose due to similar factors -- powerful but simple languages associated with effective and automated tools -- their particularities, resumed in Table~\ref{tab:compare}, make them excel in the analysis of different classes of systems. A more in-depth comparison would help the development of systems with rich static and dynamic properties, the focus of this study. For instance, Alloy users could benefit from TLC's unbounded technique, while {\TLAp} users could benefit from Alloy's flexibility. As far as we are aware, no previous study systematically compared these frameworks, so as a first step we explore the potential of embedding ``dynamic'' Alloy models in {\TLAp} using a concrete example. Although concepts are explained as needed, basic knowledge of the frameworks is expected from the reader.

\begin{table}[!t]
\centering
\scriptsize
\begin{tabular}{c|c|c}
                                        & Alloy                                            & {\TLAp} \\
              \hline
\multirow{2}{*}{\textbf{Modeling}}      & \multirow{2}{*}{Relational logic}                   & First-order logic\\
                                        &                                                     & Actions + Fairness \\
\hline
\textbf{Specification}                  & Relational logic                                    & Temporal logic \\
\hline
\multirow{2}{*}{\textbf{Verification}}  & \multirow{2}{*}{Bounded model checker (Analyzer)}& Unbounded model checker (TLC) \\
                                        &                                                     & (+ Theorem prover (TLAPS)) 
\end{tabular}
\caption{General comparison of the Alloy/{\TLAp} frameworks.}
\label{tab:compare}
\end{table}

The remainder of this paper is structured as follows. Section~\ref{sec:alloy} presents the Alloy example that will be used throughout the paper and Section~\ref{sec:tla} its embedding in {\TLAp} and our experience with TLC. Section~\ref{sec:eval} presents a comparison of the Analyzer and TLC performance for this example which are discussed in Section~\ref{sec:conc}.

\section{An Alloy Specification: Hotel Room Locking System}
\label{sec:alloy}
The \texttt{Hotel} example used throughout this paper models a hotel room locking system, initially presented by Jackson~\cite[p.~187]{Jackson:12}. This system is built on the assumption that the key management system at the front desk of the hotel is disconnected from the locking systems of the rooms. Doors unlock for the currently registered key, unless a more recent one, issued by the front desk, is detected, at which point older keys are rendered obsolete. Structural properties restrict the state of the front desk and room locking systems, while behavioral properties model how these evolve as guests check in and out. The specification will test whether unauthorized room accesses can occur. This section explores an Alloy model of this system, depicted in Fig.~\ref{fig:hotel_std}, similar to the original one apart from styling changes. For a thorough presentation of Alloy see~\cite{Jackson:12}.

\begin{figure}
\centering
\begin{alloyfig}
open util/ordering[Time] as to
open util/ordering[Key] as ko

sig Time {}

sig Key {}

sig Room {
  keys: set Key,
  current: keys one -> Time }

fact DisjointKeys {
  keys in Room lone -> Key }

one sig FD {
  last: (Room -> lone Key) -> Time,
  occupant: (Room -> Guest) -> Time }

sig Guest {
  gkeys: Key -> Time }

fun nextKey [k: Key, ks: set Key]: set Key {
  min[k.nexts & ks] }

pred Init [t: Time] {
  no Guest.gkeys.t
  no FD.occupant.t
  all r: Room | FD.last.t[r] = r.current.t }

pred Entry [t, t': Time, g: Guest, r: Room, k: Key] {
  k in g.gkeys.t
  k = r.current.t or k = nextKey[r.current.t, r.keys]
  r.current.t' = k
  all r: Room - r | r.current.t = r.current.t'
  all g: Guest | g.gkeys.t = g.gkeys.t'
  FD.last.t = FD.last.t'
  FD.occupant.t = FD.occupant.t' }

pred Checkout [t, t': Time, g: Guest] {
  some FD.occupant.t.g
  FD.occupant.t' = FD.occupant.t - Room -> g
  FD.last.t = FD.last.t'
  all r: Room | r.current.t = r.current.t'
  all g: Guest | g.gkeys.t = g.gkeys.t' }

pred Checkin [t, t': Time, g: Guest, r: Room, k: Key] {
  g.gkeys.t' = g.gkeys.t + k
  no FD.occupant.t[r]
  FD.occupant.t' = FD.occupant.t + r -> g
  FD.last.t' = FD.last.t ++ r -> k
  k = nextKey[FD.last.t[r], r.keys]
  all r: Room | r.current.t = r.current.t'
  all g: Guest - g | g.gkeys.t = g.gkeys.t' }

fact NoIntervening {
  all t: Time, t': t.next, t'': t'.next, g: Guest, r: Room, k: Key |
    Checkin[t, t', g, r, k] implies (Entry[t', t'', g, r, k] or no t'') }

fact Traces {
  Init[first]
  all t: Time, t' : t.next | some g: Guest, r: Room, k: Key |
      Entry[t, t', g, r, k] or Checkin[t, t', g, r, k] or Checkout[t, t', g] }

assert NoBadEntry {
  all t: Time, t': t.next, r: Room, g: Guest, k: Key |
    (Entry[t, t', g, r, k] and some FD.occupant.t[r]) implies g in FD.occupant.t[r] }

check NoBadEntry for 3 but 30 Time
\end{alloyfig}
\caption{Hotel room locking system under Alloy.}
\label{fig:hotel_std}
\end{figure}

\subsection{Structural and Behavioral Modeling}
\paragraph{Structure}
In Alloy, the structure of a model is defined by \emph{signatures} and their \emph{fields}. Each \texttt{Hotel} instance consists of a set of disposable keys, rooms and guests (signatures \texttt{Key}, \texttt{Room} and \texttt{Guest}, respectively). Each room has a pool of keys assigned to it (field \texttt{keys}), and keeps track of the last valid key that unlocked the door (field \texttt{current}), which must belong to the pool. A front desk (signature \texttt{FD}) keeps track of each room occupants (field \texttt{occupant}) and the last key delivered for each room (field \texttt{last}). 

Signature and field values can be further restricted by defining \emph{facts} -- constraints that must hold for every instance. In \texttt{Hotel}, fact \texttt{DisjointKeys} forces each key to belong to the pool of a single room. Finally, in order to recognize fresh keys, the locking systems and the front desk must agree on a keys order \emph{a priori}. In Alloy this can be abstracted by imposing a total order over \texttt{Key} atoms. Fresh keys are retrieved by the \texttt{nextKey} \emph{function} from the pool of available ones.

\paragraph{Behavior}
Since Alloy is inherently static, well-known idioms have been developed to model system evolution. This example follows the \emph{local state idiom}. Here, a totally ordered signature \texttt{Time} is introduced whose elements denote instants in time. Fields that should vary in time are appended with a \texttt{Time} element that denotes its value in each moment. In \texttt{Hotel} fields \texttt{current}, \texttt{last}, \texttt{occupant} and \texttt{gkeys} are expected to be variable. Although every signature is static, this is not the case in general, which would require additional facts to manage the temporal consistency of the fields.
To access values in an instant \a{t} from \texttt{Time}, one simply composes it with the expression, e.g., \a{current.t}.

Actions are specified as \emph{predicates} between two explicit instants of time (and parameters of the action). In \texttt{Hotel} the system evolves as guests check in and out and unlock the room's doors. When a guest checks in, the next key for the chosen room is given, and the guest is added to the room's occupants (predicate \texttt{Checkin}). The room's locking system is unaware of such assignment: only when a more recent key than the currently known is used to unlock the door is the system updated, rendering older keys obsolete (predicate \texttt{Entry}). Obviously, the currently known key also unlocks the door. Checking out vacates the room but allows the guest to keep the room key (predicate \texttt{Checkout}). 
To actually force the system to behave according to these predicates, fact \texttt{Traces} forces an \texttt{Init} predicate to hold in the first instant -- all rooms are vacant and no keys are assigned -- and every succeeding state to be derived from the action predicates. Now the Analyzer will only consider instances built from action application.

\subsection{Specification and Verification}
\paragraph{Specification}
The property that is expected to hold in \texttt{Hotel} is that a guest enters a room only if it is one of its occupants, a safety property. The Analyzer does not impose any restriction on the class of predicates that it is able to check. In \texttt{Hotel} this was encoded as the \texttt{NoBadEntry} \emph{assertion} -- a property that is to be checked by the Analyzer. Note how the temporal property must be explicitly defined by universally quantifying over \texttt{Time} elements (which are bound to valid evolution traces due to fact \texttt{Traces}).

\begin{figure}[t]
    \centering
    \begin{subfigure}{0.3\textwidth}
    \centering
    \includegraphics[scale=0.365]{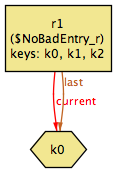}
    \caption{Initial state.}
    \end{subfigure}
    \begin{subfigure}{0.36\textwidth}
        \centering
    \includegraphics[scale=0.365]{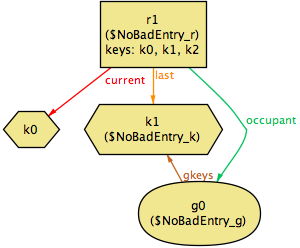}
    \caption{\a{Checkin[t,t',g0,r1,k1]}.}
    \end{subfigure}
    \begin{subfigure}{0.31\textwidth}
        \centering
    \includegraphics[scale=0.365]{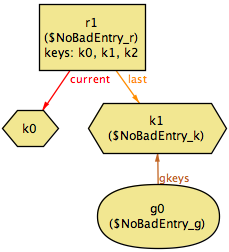}
    \caption{\a{Checkout[t,t',g0]}.}
    \end{subfigure}
    \begin{subfigure}{0.4\textwidth}
        \centering
    \includegraphics[scale=0.365]{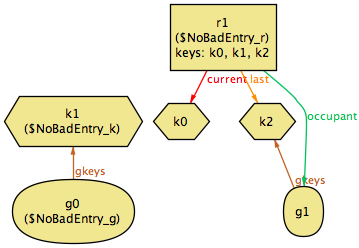}
    \caption{\a{Checkin[t,t',g1,r1,k2]}.}
    \end{subfigure}
    \begin{subfigure}{0.35\textwidth}
        \centering
    \includegraphics[scale=0.365]{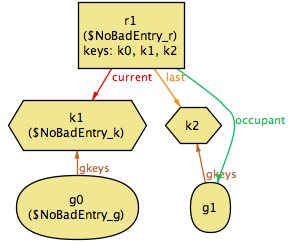}
    \caption{\a{Entry[t,t',g0,r1,k1]}.}
    \end{subfigure}
    \caption{Alloy Analyzer counter-example for \texttt{NoBadEntry}.}
    \label{fig:counter_std}
\end{figure}

\paragraph{Verification}
The Analyzer is instructed to check whether an assertion holds for a specific scope of atoms through a \emph{check} command -- here defined for 3 elements of each signature and a length trace 5. In this version of the model \texttt{NoBadEntry} does not hold, thus the Analyzer quickly generates counter-examples, one of which is depicted in Fig.~\ref{fig:counter_std} (atom names are abbreviated for readability purposes).
Since the Analyzer performs bounded model checking, the scope of \texttt{Time} imposes a limit on the length of traces, which may lead to unpredictable behaviors. The most obvious one is that counter-examples may not short traces: in \texttt{Hotel}, a scope of less than 5 instants would not detect the inconsistency, leading the user to a false sense of safety. A more subtle problem occurs when checking liveness properties, where the finite trace may lead to the generation of false positives. The turnaround is to simulate infinite traces by forcing a loop, disregarding non-infinite traces~\cite{BiereCCZ:99,Cunha:14}. These issues, allied to the need to explicit model dynamism, greatly encumber the analysis of dynamic systems in Alloy.

Another particularity of Alloy is that total orders force the number of atoms to be exactly that of the defined scope. Thus, the check command from Fig.~\ref{fig:hotel_std} only considers models with exactly 5 \texttt{Time} atoms, rather than models with \emph{up to} 5 \texttt{Time} atoms. Since there is no guarantee that counter-examples occurring in smaller traces will reappear in larger ones, the user must manually check the system for increasing \texttt{Time} scopes. Alternatively, the user could define his own notion of order that did not force exact scopes but this would have great impact on the Analyzer's performance, since the native operation is highly optimized.

The presented counter-example breaks \texttt{NoBadEntry} because, even though guest \texttt{g0} checks out of room \texttt{r1} and then guest \texttt{g1} checks in into it, the locking system of \texttt{r1} considers \texttt{g0}'s key \texttt{k1} valid until \texttt{g1} uses \texttt{k2}, rendering \texttt{k1} obsolete. A possible fix to the model is to assume that guests enter the room immediately after checking in: \texttt{g1} would immediately enter the room with \texttt{k2}, rendering \texttt{k1} invalid. Such constraint is encoded as the \texttt{NoIntervening} fact. With this fact the Analyzer no longer detects a counter-example for \texttt{NoBadEntry}.

\section{A {\TLAp} Embedding}
\label{sec:tla}
\begin{figure}[t]
\centering
\begin{subfigure}[b]{0.4\textwidth}
\centering
\scriptsize
\begin{lstlisting}[basicstyle = \ttfamily]
SPECIFICATION Spec
PROPERTY NoBadEntry
CONSTANTS KEY = 4
          GUEST = {g1,g2,g3,g4}
          ROOM = {r1,r2,r3,r4}
\end{lstlisting}
\caption{Configuration file.}
\label{fig:hotel_tlc}
\end{subfigure}
\begin{subfigure}[b]{0.55\textwidth}
    \centering
    \includegraphics[scale=0.45]{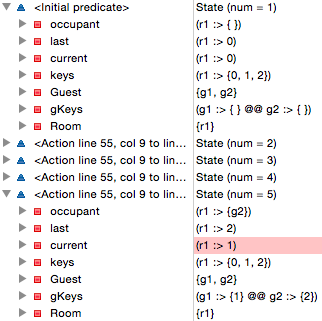}
    \caption{Counter-example for \texttt{NoBadEntry}.}
    \label{fig:counter_tlc}
\end{subfigure}
\caption{TLC configuration and counter-example for \texttt{Hotel}.}
\end{figure}

Like Alloy, {\TLAp} is an extremely expressive specification language where models consist of arbitrary predicates. However, the TLC model checker is not able to process arbitrary {\TLAp} models, imposing restrictions to the language~\cite[p.~230]{Lamport:02}. Thus, two classes of issues must be addressed throughout this embedding: first, the mismatch between the Alloy and {\TLAp} languages; second, the additional restrictions imposed by TLC for checking properties. This section thoroughly explores a possible embedding of the hotel locking system in {\TLAp}, depicted in Fig.~\ref{fig:hotel_tla}. This specification is TLC-compatible, and can be deployed under the configuration%
\footnote{%
The TLA tools are currently deployed as the \emph{TLA Toolbox}, where the TLC parameters are defined through the GUI rather than as a configuration file.} %
presented in Fig.~\ref{fig:hotel_tlc}. Although {\TLAp} concepts are presented as required, the interested reader is redirected to~\cite{Lamport:02} for an in-depth presentation.

\begin{figure}
\tlatex
\footnotesize
\fl{}\moduleLeftDash\cl{ {\MODULE} Hotel}\moduleRightDash\cl{}
\fl{ {\EXTENDS} Naturals}
\fl{ {\CONSTANT} KEY , ROOM , GUEST}
\fl{ {\ASSUME} KEY \.{\in} Nat}
 \fl{ {\VARIABLE} keys , current , last , occupant , gkeys , Room
 , Guest}
\fl{}\midbar\cl{}
\fl{ Key \.{\defeq} 0 \.{\dotdot} KEY \.{-} 1}
\fl{}
 \fl{ TypeInv \.{\defeq}}\al{9}{2}{ \.{\land}}\al{9}{3}{ Room \.{\in}
 {\SUBSET} ROOM \.{\land}}\al{10}{1}{ Guest \.{\in} {\SUBSET} GUEST}
 \fl{\qquad\qquad\quad\, \.{\land}}\al{11}{1}{ keys \.{\in} [ Room \.{\rightarrow} {\SUBSET} Key
 ] \.{\land}}\al{15}{1}{ last \.{\in} [ Room \.{\rightarrow} Key ]}
\fl{ \qquad\qquad\quad\, \.{\land}}\al{12}{1}{ current \.{\in} [ Room \.{\rightarrow} Key ] \.{\land}}\al{16}{1}{ occupant \.{\in} [ Room \.{\rightarrow} {\SUBSET}
 Guest ]}
 \fl{ \qquad\qquad\quad\,\.{\land}}\al{17}{1}{ gkeys \.{\in} [ Guest \.{\rightarrow} {\SUBSET}
 Key ]}
 \fl{ \qquad\qquad\quad\,\.{\land}}\al{13}{1}{ \A\, r \.{\in} Room \.{:} current [ r ] \.{\in}
 keys [ r ]}
 \fl{ \qquad\qquad\quad\,\.{\land}}\al{14}{1}{ \A\, r1 , r2 \.{\in} Room \.{:} ( keys [ r1 ]
 \.{\cap} keys [ r2 ] ) \.{\neq} \{ \} \.{\impliesT} r1 \.{=} r2}
\fl{}
 \fl{ Init \.{\defeq}}\al{19}{2}{ \.{\land}}\al{19}{3}{ Room \.{\in} {\SUBSET}
 ROOM \.{\land}}\al{20}{1}{ Guest \.{\in} {\SUBSET} GUEST}
 \fl{ \qquad\quad\,\,\, \.{\land}}\al{21}{1}{ keys \.{\in} [ Room \.{\rightarrow} {\SUBSET} Key
 ] \.{\land}}\al{27}{1}{ gkeys \.{=} [ g \.{\in} Guest \.{\mapsto} \{ \} ] }
\fl{ \qquad\quad\,\,\, \.{\land}}\al{22}{1}{ current \.{\in} [ Room \.{\rightarrow} Key ] \.{\land}}\al{26}{1}{ occupant \.{=} [ r \.{\in} Room \.{\mapsto} \{ \}
 ]}
 \fl{ \qquad\quad\,\,\, \.{\land}}\al{23}{1}{ \A\, r \.{\in} Room \.{:} current [ r ] \.{\in}
 keys [ r ]}
 \fl{ \qquad\quad\,\,\, \.{\land}}\al{24}{1}{ \A\, r1 , r2 \.{\in} Room \.{:} ( keys [ r1 ]
 \.{\cap} keys [ r2 ] ) \.{\neq} \{ \} \.{\impliesT} r1 \.{=} r2}
\fl{ \qquad\quad\,\,\, \.{\land}}\al{25}{1}{ last \.{=} current}
\fl{}
 \fl{ vs \.{\defeq} {\langle} keys , current , last , occupant , gkeys
 , Guest , Room {\rangle}}
\fl{}
 \fl{ nextKey [ k \.{\in} Key , ks \.{\in} {\SUBSET} Key ] \.{\defeq} \{ x
 \.{\in} ks \.{:} x \.{>} k \.{\land} ( \A\, y \.{\in} ks \.{:} y \.{>} k
 \.{\impliesT} x \.{\leq} y ) \}}
\fl{}
 \fl{ Entry ( g , r , k ) \.{\defeq}}\al{33}{9}{ \.{\land}}\al{33}{10}{ k
 \.{\in} gkeys [ g ]}
 \fl{ \qquad\qquad\qquad\quad\,\,\,\, \.{\land}}\al{34}{1}{ ( k \.{=} current [ r ] \.{\lor} \{ k \} \.{=}
 nextKey [ current [ r ] , keys [ r ] ] )}
 \fl{ \qquad\qquad\qquad\quad\,\,\,\, \.{\land}}\al{35}{1}{ current \.{'} \.{=} [ current {\EXCEPT} {\bang} [
 r ] \.{=} k ]}
 \fl{ \qquad\qquad\qquad\quad\,\,\,\, \.{\land}}\al{36}{1}{ {\UNCHANGED} {\langle} keys , last , occupant
 , gkeys , Guest , Room {\rangle}}
\fl{}
 \fl{ Checkout ( g ) \.{\defeq}}\al{38}{5}{ \.{\land}}\al{38}{6}{ \E\, r
 \.{\in} Room \.{:} g \.{\in} occupant [ r ]}
 \fl{ \qquad\qquad\qquad\quad\, \.{\land}}\al{39}{1}{ occupant \.{'} \.{=} [ r \.{\in} {\DOMAIN}
 occupant \.{\mapsto} occupant [ r ] \.{\,\backslash\,} \{ g \} ]}
 \fl{ \qquad\qquad\qquad\quad\,\.{\land}}\al{40}{1}{ {\UNCHANGED} {\langle} keys , last , current
 , gkeys , Guest , Room {\rangle}}
\fl{}
 \fl{ Checkin ( g , r , k ) \.{\defeq}}\al{42}{9}{ \.{\land}}\al{42}{10}{
 occupant [ r ] \.{=} \{ \} \.{\land}}\al{43}{1}{ \{ k \} \.{=} nextKey [ last [ r ] , keys [ r ]
 ]}
 \fl{ \qquad\qquad\qquad\qquad\,\,\, \.{\land}}\al{44}{1}{ occupant \.{'} \.{=} [ occupant {\EXCEPT} {\bang}
 [ r ] \.{=} \{ g \} ]}
 \fl{ \qquad\qquad\qquad\qquad\,\,\, \.{\land}}\al{45}{1}{ gkeys \.{'} \.{=} [ gkeys {\EXCEPT} {\bang} [ g ]
 \.{=} @ \.{\cup} \{ k \} ]}
 \fl{ \qquad\qquad\qquad\qquad\,\,\, \.{\land}}\al{46}{1}{ last \.{'} \.{=} [ last {\EXCEPT} {\bang} [ r ]
 \.{=} k ]}
 \fl{ \qquad\qquad\qquad\qquad\,\,\, \.{\land}}\al{47}{1}{ {\UNCHANGED} {\langle} keys , current , Guest
 , Room {\rangle}}
\fl{}
 \fl{ Post ( g , r , k ) \.{\defeq}  occupant [ r ] \.{=} \{ g \}  \.{\land}}\al{50}{1}{ k \.{\in} gkeys [ g ] \.{\land}}\al{51}{1}{ last [ r ] \.{=} k \.{\land}}\al{52}{1}{ current [ r ] \.{\neq} k%
}
\fl{}
 \fl{ Act \.{\defeq} \E\, g \.{\in} Guest \.{:} Checkout ( g ) \.{\lor} \E\,
 r}\al{54}{14}{ \.{\in}}\al{54}{15}{ Room , k \.{\in} Key \.{:} Entry ( g
 , r , k ) \.{\lor} Checkin ( g , r , k )}
\fl{}
 \fl{ NoIntervening \.{\defeq} \A\, g \.{\in} Guest , k \.{\in} Key ,
 r}\al{56}{12}{ \.{\in}}\al{56}{13}{ Room \.{:} Post ( g , r , k )
 \.{\impliesT} Entry ( g , r , k )}
 \fl{}
 \fl{ Spec \.{\defeq} Init \.{\land} {\Box} [ NoIntervening \.{\land} Act
 \.{\land} TypeInv ]_{ vs}}
\fl{}\midbar\cl{}
\fl{ NoBadEntry \.{\defeq} {\Box} [ \A\, g \.{\in} Guest , r \.{\in} Room , k \.{\in} Key \.{:}}
 \fl{ \qquad\qquad\qquad\qquad\qquad\qquad
 Entry ( g , r , k ) \.{\land} occupant [ r ] \.{\neq} \{ \} \.{\impliesT}
 g \.{\in} occupant [ r ] ]_{ vs}}
\fl{}\bottombar\cl{}%
\caption{The {\TLAp} embedding of the hotel locking system.}
\label{fig:hotel_tla}
\end{figure}

\subsection{Structural and Behavioral Modeling}
\paragraph{Structure}
Alloy signatures and fields take the shape of module parameters in {\TLAp}, which can be either \emph{constant} or \emph{variable}. This differs from the object-oriented flavor of Alloy, where fields must be associated with a parent signature. As a consequence, \texttt{current} and \texttt{occupant} no longer need to belong to the placeholder object \texttt{FD}, which is discarded in the {\TLAp} version. 

The user could expect to represent the static components of the model (sets \texttt{Key}, \texttt{Room} and \texttt{Guest} and relation \texttt{keys}) as constant parameters. However, to deploy TLC, the user must assign a fixed value to constant parameters. This differs from the behavior of static components in the Alloy version, leading to unpredictable results: although fields like \texttt{keys} do not change over time, the Analyzer explores the evolution of the system under every valid assignment to those variables. 
A similar issue occurs with \texttt{Room} and \texttt{Guest} since the scope \a{for 3} only sets the maximum number of atoms per signature (unless the user imposes an exact scope or a total order). Defining them as constant in {\TLAp} and assigning them 3 elements in the TLC configuration would assign them \emph{exactly} those elements. This is problematic in \texttt{Hotel} because such configurations do not generate counter-examples. The model forces at least one key to be assigned to each room; the instance from Fig.~\ref{fig:counter_std} arises from the successive checking in of two guests in the same room, requiring a universe with two extra keys: a distracted user could specify \texttt{Key = \{1,2,3\}} and \texttt{Room = \{r1,r2,r3\}} and miss it. 

To model the expected behavior, \texttt{keys}, \texttt{Guest} and \texttt{Room} are specified as variable parameters, the last two being forced to be within constants \texttt{GUEST} and \texttt{ROOM} denoting their universe. Their values are unchanged by the actions, so TLC will assign them an arbitrary value in the initial state but preserve it throughout the trace, simulating the semantics of the Alloy version. 

{\TLAp} is untyped, but it is considered good practice to declare a type invariant over the variable parameters that is expected to hold in every state~\cite{Lamport:02}. These are defined as \emph{state predicates} -- first-order logic formulas without temporal operators -- as \texttt{TypeInv} in Fig.~\ref{fig:hotel_tla}. We opted to encode the variables corresponding to Alloy fields as functions, because these are more manageable in {\TLAp} and the \texttt{Hotel} model does not rely on relational operators. {\TLAp} does not support relational operations, like the converse or transitive closure, natively, so converting other Alloy models may not be as straightforward. 
Expression ${\SUBSET} A$ denotes the power-set of $A$, and is used to ``type'' many-valued functions. For simplicity, \texttt{last} was implemented as a total function from rooms to keys (from \texttt{Init} and the defined actions, it would be always defined for every room). Additional structural constrains, like the fact that the \texttt{keys} assignments are disjoint and that the \texttt{current} key is selected from these pools, are also defined in \texttt{TypeInv}. 

Since constant parameters are bound in the TLC configuration, {\TLAp} instead provides an $\ASSUME$ keyword that should be used to restrict their values, which do not affect the meaning of the model but instruct TLC to test whether the assignments of the constant parameters are valid. 
The \texttt{Key} set should be totally ordered and bound exactly, so a constant \texttt{KEY}, assumed to be a natural, sets the number of keys available, from which predicate \texttt{Key} creates a range. The \texttt{nextKey} function then relies on natural intervals to retrieve the next key.

Likewise Alloy, defining a predicate like \texttt{TypeInv} does not affect the model by itself. Nonetheless, the expected role of type invariants in Alloy and {\TLAp} renders evident a difference in methodology. In {\TLAp}, these are typically not enforced, but instead used to check the correctness of the defined actions~\cite[p.~26]{Lamport:02}. This reflects the prominent role of the actions in {\TLAp}, that, along with the initial state predicate, entail the set of acceptable states. In fact, TLC-compatible actions must completely specify the succeeding states~\cite[p.~238]{Lamport:02}. In contrast, actions in Alloy are defined by regular declarative predicates and do not by themselves entail the set of valid states: type invariants imposed by facts may restrict their result. This allows the user to separate the concerns between the structural and behavioral components of the model, and obtain simpler action definitions. In order to be as faithful to the Alloy model, \texttt{TypeInv} will be forced%
\footnote{%
We also tested \texttt{Hotel} in TLC with \texttt{TypeInv} set as an invariant, which proved that the defined actions effectively preserve the type invariant.
}%
.


\paragraph{Behavior}
A model in {\TLAp} is expected to follow the shape $Init \wedge {\Box} [Next]_vs \wedge Temporal$, where $Init$ is a state predicate restricting the initial state, $Next$ restricts valid evolution steps through \emph{actions} -- predicates with primed variables referring to their value in the succeeding state -- and $Temporal$ is an additional temporal restriction. Stuttering steps are intrinsic in {\TLAp}, $[Next]_vs$ denoting that either $Next$ holds or a stuttering step is performed with $v' = v$. TLC only supports fairness restrictions in $Temporal$, which do not exist in \texttt{Hotel}.

Predicate \texttt{Init} follows the semantics of the corresponding Alloy predicate, where \texttt{occupant} and \texttt{gkeys} are assumed to be empty, and \texttt{last} to have the same value as \texttt{current}. As for \texttt{keys}, it is free to take any value that results in a disjoint set, from which the \texttt{current} key is selected. 
Predicate \texttt{Act} is comprised by the disjunction of the \texttt{Entry}, \texttt{Checkin} and \texttt{Checkout} actions. This is similar to the Alloy approach, except that dynamism is now intrinsic to the language through primed variables. {\TLAp} provides some syntactic sugar to define frame conditions: an ${\UNCHANGED} x$ expression is an abbreviation for $x' = x$ and for sequences or functions, an expression $f' = [f {\EXCEPT} ![x] = e]$ means that $f$ remains unchanged except for its value at $x$, which is updated to $e$ ($@$ denotes the previous $f[x]$ value). Although the model from Fig.~\ref{fig:hotel_tla} relies on these, it is not clear whether they can be derived from an Alloy model. Predicate \texttt{TypeInv} is also forced in every state by being introduced in $Next$.  Alternatively it could be set as a \texttt{CONSTRAINT} in the TLC configuration, which would ignore states where it does not hold, but the model would not be self-contained.

In order to be TLC-compatible however, these predicates must follow some stricter rules that bound the possible values of the parameters. For instance, when a parameter is referenced its value must have already been bound by a previous conjunct. Thus, in \texttt{Init}, expression $\forall r \in Room : current[r] \in keys[r]$ cannot occur before $keys \.{\in} [ Room \.{\rightarrow} {\SUBSET} Key ]$ and $current \.{\in} [ Room \.{\rightarrow} Key ]$, which define the upper bounds of \texttt{keys} and \texttt{current}.
TLC must also be able to derive the value of every parameter in the succeeding state from $Next$. This contrasts with the definition of actions in Alloy that are purely declarative predicates, which may not bound the next state completely: either their outputs are restricted by other facts or their behavior is non-deterministic. Thus, it is not clear whether TLC-compatible actions can be derived from Alloy predicates. This hints again at the central role of actions in {\TLAp} rather than structural properties.

\subsection{Specification and Verification}
\paragraph{Specification}
Unlike Alloy, {\TLAp} supports the specification of a subset of LTL properties, supporting the \emph{always} ${\Box}$ and \emph{eventually} ${\Diamond}$ temporal quantifiers. However, the class of formulas that TLC is able to check is limited~\cite[p.~236]{Lamport:02}: they must either be state predicates $P$, invariant predicates ${\Box} P$, box-action formulas ${\Box}[A]_vs$ or simple temporal formulas, i.e., boolean combinations of temporal state formulas (${\Box}P$, ${\Diamond}P$ or ${\Box}(P {\impliesT} {\Diamond}Q)$) and simple action formulas (expressions $\Box\Diamond\langle A \rangle_vs$, $\Diamond\Box[A]_vs$ and fairness predicates). This contrasts with Alloy where any predicate supported by the language can be checked by the Analyzer. Thus, to check the validity of \texttt{NoBadEntry}, the predicate must be converted into a valid {\TLAp} formula that TLC is able to process. 
Luckily, \texttt{NoBadEntry} can be translated into a TLC compatible {\TLAp} formula in a straightforward manner in the shape of a box-action formula $\Box[A]_vs$, as defined in Fig.~\ref{fig:hotel_tla}.
TLC is instructed to check properties through \texttt{PROPERTY} instructions in the configuration, as introduced in Fig.~\ref{fig:hotel_tlc}. 
Defining \texttt{NoBadEntry} as a box-action raises some issues related with {\TLAp}'s assumption of stuttering steps. When a guest enters the room for the second time -- i.e., when his key has already been registered -- the \texttt{Entry} action does not update the value of any variable. Thus, TLC will not recognize such steps, identifying instead a stuttering step.


\paragraph{Verification}
By default, TLC will flag deadlocks as errors. The \texttt{Hotel} model is however expected to enter deadlock for configurations without enough keys, and due to the identification of \texttt{Entry} steps as stuttering steps as explained above, deadlocks occur in every configuration. Thus, TLC must be instructed to ignore deadlocks. As expected, TLC will then find a counter-example for \texttt{NoBadEntry}, which is depicted in Fig.~\ref{fig:counter_tlc}.

To guarantee that \texttt{NoBadEntry} holds, the behavioral model must be further restricted. Unfortunately, \texttt{NoIntervening} is not a valid formula in the {\TLAp} language because variables cannot be doubly primed. This renders the expression \a{Entry[t',t",g,r,k]} not expressible in {\TLAp}.
Thus, \texttt{NoIntervening} was adapted to force an \texttt{Entry} action to occur whenever there is a guest whose key is not registered in the room, which is the post-condition of the \texttt{Checkin} action, as defined by \texttt{Post} in Fig.~\ref{fig:hotel_tla}.
Such predicate is added to $Next$, forbidding invalid state transitions. Likewise \texttt{CONSTRAINT} instructions, \texttt{ACTION-CONSTRAINT} instructions could also have been used. Under this configuration, TLC does not find any counter-example for \texttt{NoBadEntry}, as expected.
One could wonder whether it would not suffice to check $NoIntervening \impliesT NoBadEntry$. Unfortunately, this is not a valid TLC specification, since it is not a simple temporal formula.

\section{Evaluation}
\label{sec:eval}

The Analyzer and TLC check properties through fundamentally different techniques. The Analyzer embeds models into propositional formulas which are fed to off-the-shelf SAT solvers. Thus, the procedure is oblivious of the model that originated the formula. This also renders the process bounded, so temporal properties may only be checked for traces with limited length.
In contrast, TLC interprets {\TLAp} models as finite state machines and deploys an explicit-state model checker. Thus, it considers traces with unlimited length and has a finer control on how the states are explored, enabling breadth-first searches.

This section compares the performance of the two approaches in the verification of the \texttt{Hotel} model. All tests were performed multiple times on an 1,8 GHz Intel Core i5 with 4 GB memory running OS X 10.10 with Alloy Analyzer 4.2 and TLC 2.08.
We are interested in assessing how the existence of counter-examples affects their performance, so tests were run with and without the \texttt{NoIntervening} constraint enforced. Although TLC's default model is breadth-first (BF), it also supports depth-first (DF) searches, which was also tested (for a maximum depth of 100) since it could fare better when there are counter-examples to be found. TLC was also instructed not to flag deadlocks as errors. The Analyzer was run with the \textsf{MiniSat} solver, which our experiments show to be the most efficient. 

\begin{figure}[t]
    \centering
    \begin{subfigure}{0.495\textwidth}
    \centering
    \includegraphics[scale=0.375]{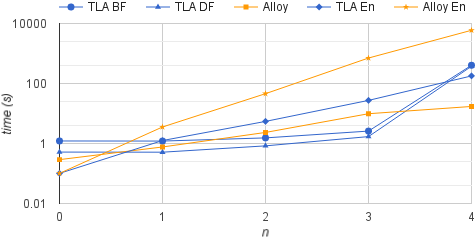}
    \caption{Counter-example, $t=30$.}
    \label{fig:t1}
    \end{subfigure}
    \begin{subfigure}{0.495\textwidth}
    \centering
    \includegraphics[scale=0.375]{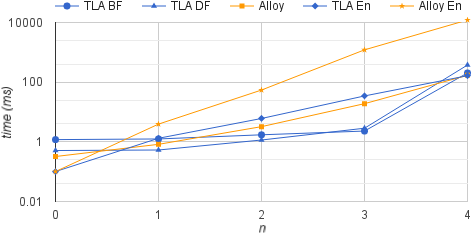}
    \caption{\texttt{NoIntervening}, $t=30$.}
    \label{fig:t2}
    \end{subfigure}
    \begin{subfigure}{0.495\textwidth}
    \centering
    \includegraphics[scale=0.375]{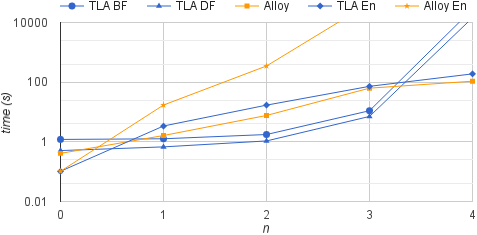}
    \caption{Exact scope, counter-example, $t=30$.}
    \label{fig:e1}
    \end{subfigure}
    \begin{subfigure}{0.495\textwidth}
    \centering
    \includegraphics[scale=0.375]{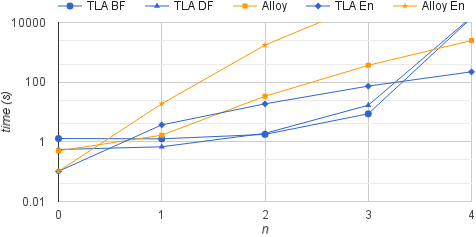}
    \caption{Exact scope, \texttt{NoIntervening}, $t=30$.}
    \label{fig:e2}
    \end{subfigure}
    \centering
    \begin{subfigure}{0.495\textwidth}
    \centering
    \includegraphics[scale=0.375]{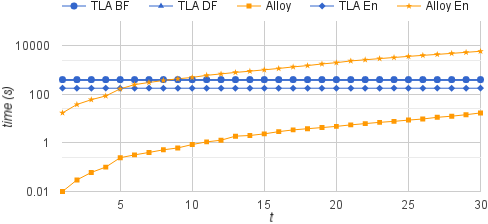}
    \caption{Counter-example, $n=4$.}
    \label{fig:n1}
    \end{subfigure}
    \begin{subfigure}{0.495\textwidth}
    \centering
    \includegraphics[scale=0.375]{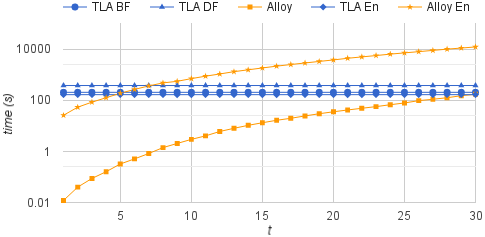}
    \caption{\texttt{NoIntervening}, $n=4$.}
    \label{fig:n2}
    \end{subfigure}
    \caption{\texttt{Hotel} performance under different approaches.}
    \label{fig:performance}
\end{figure}

The model was tested for different $n$ sizes, denoting the number of keys and the maximum number of guests and rooms. 
Besides being bounded, the total ordering in Alloy forces the trace length to be fixed, so tests are run up to $t$ in order to find counter-examples with minimal trace length: the timing for a given $t$ aggregates the timing of the previous $t-1$ runs. TLC's breadth-first searches naturally find such counter-examples. The check command from Fig.~\ref{fig:hotel_std} and the configuration from Fig.~\ref{fig:hotel_tlc} represent the model for $n = 4$ (and $t=30$ for Alloy).
Figures~\ref{fig:t1} and~\ref{fig:t2} compare the performance of the approaches for $t=30$ (relevant only for the Alloy results) and increasing size $n$. In general, although TLC fares better for smaller $n$ values, it is outperformed by the Analyzer for larger ones. Interestingly, TLC is actually faster with \texttt{NoIntervening} enabled than without it. 
Figures~\ref{fig:n1} and~\ref{fig:n2} compare instead the performance of the approaches for a fixed $n=4$ with increasing trace length $t$ (the unbounded TLC results appear as a constant function on $t$). As seen above, the Analyzer outperforms TLC for such $n$ values even at $t=30$, e.g., the Analyzer detects the counter-example at $t=5$ in less than a second in contrast to TLC's 415 seconds.

Our experiments show that TLC spends a great deal of time generating every initial state as a first step. For $n=3$ and $n=4$, there are 776 and 18960 distinct initial states, respectively, arising from the attribution of values to \texttt{Guest}, \texttt{Room}, \texttt{keys} and \texttt{current}. Even with \texttt{NoIntervening} disabled, although TLC detected rather quickly that there is a counter-example, it spends considerable time trying to reconstruct it from the trace, which also seems to be related to the number of initial states. Interestingly, this also occurs in depth-first mode. 
In some contexts, this procedure could pay off due to its ability to detect equivalent states. However, in \texttt{Hotel} (and in fact, in problems with rich structural in general), state sharing is limited since the free variable parameters in the initial state remain fixed as the system evolves, not overlapping with other traces. 

To compare the performance of the techniques in a scenario with less initial states, we modeled a version of \texttt{Hotel} with \emph{exactly} $n$ \texttt{Guest} and \texttt{Room} elements and $n+2$ \texttt{Key} elements (the two scenarios do not have the same universe for the same $n$ value and thus are not directly comparable, but without extra keys counter-examples would not occur). For $n=3$ and $n=4$ there are now 108 and 960 initial states, respectively. The results are presented in Figs.~\ref{fig:e1} and~\ref{fig:e2}. Results show that although TLC is still outperformed by the Analyzer.

To overcome TLC's generation of initial states we explore a different approach, where the Analyzer is used to generate them, and then TLC is deployed over each fixed initial state. This way the weight of solving rich structural constraints is shifted to the more efficient Alloy, while TLC analyzes only the temporal properties as it is designed to do. Moreover, the problem can be partitioned into as many problems as there are initial states, avoiding executions running out of memory. The trade-off is that counter-examples will no longer be guaranteed to have minimal trace lengths since initial states are generated  in an arbitrary order. Also, TLC will no longer be able to benefit from state sharing.
An Alloy model that generates such initial states can be derived from the dynamic one by removing every constraint that refers to instants of time other than the first, and then, removing every reference to the \texttt{Time} signature (adapting the multiplicities of the fields accordingly). 
An additional benefit is that Alloy's symmetry breaking greatly reduces the number of generated states -- from 18960 to 520 for $n=4$, which are generated in less than 2 seconds (with symmetry breaking disabled, it produces the 18960 states). TLC also allows the definition of the model elements as symmetric, but in \texttt{Hotel} this does not seem to affect the number of states.

%
%
%
%
%
%
%
%

For a preliminary study, we timed the generation of these initial states by Alloy and the execution of TLC and the Analyzer for a set of fixed initial states. The performance of this technique when there are counter-examples to be found is dependent on how many initial lead to a the counter-example. We calculated this ratio and the times with \texttt{NoIntervening} disabled consider only considered a portion of the initial states. With \texttt{NoIntervening} enforced, all initial states were considered. The results regarding this two-phased technique are presented under ``TLC En'' in and ``Alloy En'' in Fig.~\ref{fig:performance}.
TLC with fixed initial states seems to outperform regular TLC executions, in some cases quite significantly, although its gains over pure Alloy are not so evident. Alloy with fixed initial states fares worse than all the other techniques, which is expected since Alloy has poor support for partial instances. 


\section{Discussion and Future Work}
\label{sec:conc}

Throughout this paper some pros and cons of using Alloy and {\TLAp} to specify systems with rich static and dynamic properties were identified. Alloy has two main limitations: first, dynamism must be explicitly modeled by the user, which is a cumbersome and error-prone task, even if following well-known idioms; second, the Analyzer only supports bounded model checking, which hinders the sound verification of temporal properties. The fact that properties are verified for exact trace lengths renders this process even more cumbersome.
Alloy's major advantage lies in its expressiveness, fully supported by the Analyzer, in contrast to the restrictions imposed by TLC over {\TLAp} models. This is patent in the definition of the initial state and the action predicates, which in Alloy may be purely declarative predicates but that must completely specify the state in {\TLAp} to be processed by TLC. The properties that TLC is able to verify are also restricted. The explicit dynamism in Alloy actually ends up being more expressive than {\TLAp}, as the free use of doubly primed variables demonstrate. Finally, the management of non-variable arbitrary artifacts, common in problems with rich structural properties are better manageable in Alloy than in {\TLAp}, where constants must be assigned exact values.

Regarding the possible embedding of Alloy models and specifications into {\TLAp}, some mismatches have been identified that would hinder this process. First, the translation of predicates rich in relational operators would not be straight-forward. This would be especially problematic with transitive closure operations, that must be converted to recursive definitions in {\TLAp}. Second, it is not clear whether deriving TLC compatible models (initial state and action predicates) and specifications to be verified from arbitrary Alloy predicates would be possible. It is also not clear how enforcing stuttering steps in {\TLAp} would affect the semantics of actions derived from Alloy.

As for performance, the Analyzer seems to outperform TLC with larger scopes, but with the obvious caveat of the bounded search space. Feeding {\TLAp{}} initial states generated by Alloy showed promising results, but would depend on an effective translation between the two languages. We expect to explore such technique in the future, namely by relying directly on Alloy's underlying model checker Kodkod~\cite{Torlak:07}, which has native support for partial instances.


\bibliographystyle{abbrv}
\bibliography{temporal} 

\end{document}